\documentclass[runningheads]{CMSIM}

\usepackage{graphicx} 
\usepackage{amsmath,amssymb}                      
\usepackage{natbib}
\usepackage{subfigure}
\usepackage{caption}
\usepackage{romannum}

\renewcommand{\thefootnote}

\title*{Diagnostics and Visualization of Point Process Models for Event Times on a Social Network}

\titlerunning{\it Model Checking for Network Point Processes}

\author{
Jing Wu\inst{1},
Anna L. Smith\inst{1}
\and
Tian Zheng\inst{1}
}

\authorrunning{\it Wu et al.}

\institute{
1. Department of Statistics, Columbia University, New York, New York, USA\\
* Corresponding Author: Tian Zheng, E-mail: {\tt tian.zheng@columbia.edu}
}

\begin{document}
\thispagestyle{empty}
\maketitle             
\setlength{\leftskip}{0pt}
\setlength{\headsep}{16pt}
\begin{abstract}
Point process models have been used to analyze interaction event times on a social network, in the hope to provides valuable insights for social science research. However, the diagnostics and visualization of the modeling results from such an analysis have received limited discussion in the literature. In this paper, we develop a systematic set of diagnostic tools and visualizations for point process models fitted to data from a network setting. We analyze the residual process and Pearson residual on the network by inspecting their structure and clustering structure. Equipped with these tools, we can validate whether a model adequately captures the temporal and/or network structures in the observed data. The utility of our approach is demonstrated using simulation studies and point process models applied to a study of animal social interactions.
\end{abstract}

\section{Introduction}
\begin{figure}[h]
 \centering
 \subfigure
 {
 \includegraphics[width=0.47\linewidth]{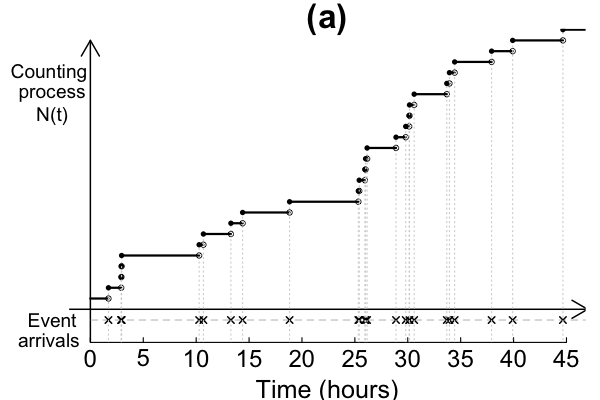}
 }
 \subfigure
 {
 \includegraphics[width=0.47\linewidth]{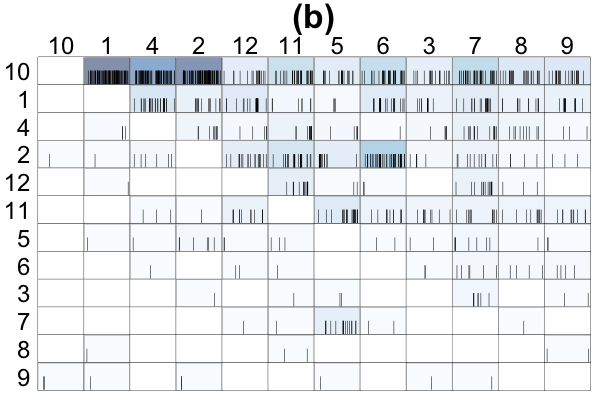}
 }
 \caption{Interaction event times on a social network of mice from \citet{williamson2016temporal}. (a) The univariate point process between one pair of mice. The crosses on the bottom indicating the observed event times. The counting process $N(t)$, defined as {\em the number of events observed up to time $t$}, is plotted. (b) Network point processes. Each row represents the initiator (sender) and each column indicates the recipient (receiver) for a social interaction. The rows (columns) are reordered by the I\&SI rank \citep{schmid2013finding}, where the top row (left column) is the most dominant mouse. The I\&SI ranking is a widely used dominance ranking method from animal behavior literature. For the square at $i$-th row and $j$-th column, the observed event times from $i$ to $j$ are plotted as line segments at the bottom. The color shade reflects the total number of events for each pair, with darker color representing more events.}
 \label{fig:real_example}
\end{figure}

Interaction event times observed on a social network provide valuable information for social scientists to gain insight into the dynamics and dependence structure among actors on this network. \citet{williamson2016temporal} studies the social interaction patterns of group-housed male mice over long time periods. In Figure~\ref{fig:real_example}-(a), we plot the univariate event times from one pair of mice, which show irregular event-intense intervals and heterogeneity in event densities. Figure~\ref{fig:real_example}-(b) displays the observed social interactions among one cohort of mice over time. The interaction patterns appear to be heterogeneous and structured across the social network. We will apply the proposed diagnostic tools to analyses of this data set as real data examples throughout this paper.

A number of continuous-time social network event times models have been recently developed. \citet{saito2009learning} studies the cross-reference network of blogs and proposes a model for topic propagation diffusion dynamics. 
\citet{fan2009learning} considers a continuous-time Bayesian network with time-varying nodal attributes. Point process models with network dependence structure have become common practice for such studies, with a focus on intensity function estimation \citep[e.g.,][]{perry2013point,linderman2014discovering,zipkin2016point,yang2017decoupling}. In Section \ref{sec:background}, we provide a short review of point process models and existing strategies for incorporating network structure. 

Despite the amount of effort that has been devoted to developing such models, there is limited work on the assessment and diagnostics of those models. To summarize and compare a model's performance, the likelihood of the model with respect to the observed data is usually evaluated. As an overall assessment of model fit, however, it does not provide detail for detecting \textit{when}, \textit{where}, and \textit{how} the proposed model inadequately accounts for patterns and variations in the observed data. Especially, for interaction dynamics on a social network, the timing often exhibits a bursty and heavy-tailed pattern \citep{barabasi2005origin}, while the structure of interdependence among network actors can be heterogeneous and complicated. Further model developments for addressing these challenges call for the development of better tools that allow researchers to systematically examine and identify the lack-of-fit of existing models against observed data.

The goal of this paper is to propose diagnostic statistics and visualization tools for network event times models, which we develop as extensions of evaluation techniques for univariate point process models. We investigate the proposed techniques using simulated studies and real data. In Section \ref{sec:background}, we introduce the notation for point processes, network point processes, and related models. Section \ref{sec:time} focuses on time-domain diagnoses by applying the time rescaling theorem and inspecting residual processes. The diagnostic tools for detecting network heterogeneity and network structure in residual processes are developed and demonstrated in Section \ref{sec:network}. 

\section{Background}
\label{sec:background}
\subsection{Univariate point processes}

Let $(0,T]$ be the time interval of observation. We denote the history of arrival times of observed events up to $T$ as $\mathcal{H}(T) = \{t_m\}_{m=0}^M$, where $t_0=0$, $t_M=T$, and $M$ is the total number of events. Figure \ref{fig:real_example}-(a) is a point process of interaction event times between one pair of mice from the animal behavior study in \cite{williamson2016temporal}.
The associated univariate point process is defined via a {\em counting process}, $N(t)$, $t\in(0,T]$, where $N(t)$ is a right-continuous function that records the number of events observed during the interval $(0,t]$. The stochastic properties of a point process is usually specified by its conditional intensity function $\lambda(t|\mathcal{H}(t))$ at any time $t$, $$\lambda(t|\mathcal{H}(t))=\lim_{\Delta t\to 0} \frac{Pr([N(t+\Delta t)-N(t)]=1|\mathcal{H}(t))}{\Delta t}.$$ Inference on the intensity function is conducted by evaluating the likelihood function, $$\prod_{m=1}^{M}\lambda(t_m|\mathcal{H}(t_m))\exp\Big\{-\int_0^T\lambda(s|\mathcal{H}(s))ds\Big\}.$$

A \textit{homogeneous Poisson process} is the simplest model and assumes a constant intensity $\lambda(t)\equiv \lambda, \lambda>0$. It cannot accommodate situations where event densities vary as shown in Figure~\ref{fig:real_example}-(a). A \textit{Hawkes process} \citep{hawkes1971spectra} is a self-exciting process whose intensity has the form $$\lambda(t) = \lambda_1 + \sum_{t_m<t}\phi_{\theta}(t-t_m),$$ where $\phi_{\theta}(t)$ is a self-exciting kernel with parameter $\theta$. The exponential kernel $\phi(t)=\alpha\exp{(-\beta t)}$ is most widely-used, where $\alpha>0$ calibrates the instantaneous boost to the event intensity at each arrival of an event, and $\beta>0$ controls the decay of past events' influence over time. 

When event dynamics display a bimodal pattern, for example, an alternation between long waiting times and intervals of more intensive events, some models assume that the intensity is modulated by a latent continuous-time Markov chain, $Z(t)$. Here, a $Z(t)$ with $S$-states is parameterized by its infinitesimal generator matrix $Q\in\mathbb{R}^{S\times S}$ \citep{rabiner1989tutorial}. To accommodate these bimodal patterns, \cite{fischer1993markov} proposed a \textit{Markov-modulated Poisson Process} (MMPP) model whose intensity function is $\lambda_{Z(t)}$. When the Markov process $Z(t)$ is in state $s\ (s=1,...,S)$, arrivals occur according to a homogeneous Poisson process of rate $\lambda_s$. Instead of simply using a constant rate $\lambda_s$ in the MMPP, \citet{wang2012markov} assumes Hawkes processes with piecewise constant kernel functions as \textit{Markov-modulated Hawkes processes with stepwise decay} (MMHPSD). The inference procedure heavily relies on the piecewise constant assumption. As a result, MMHPSD suffers from the problem that the inferred latent state is highly sensitive to single events and hard to interpret. \cite{zheng_markov-modulated_2019} introduces the more widely-used exponential kernel for $\lambda_s(t)$ in the \textit{Markov-modulated Hawkes Process} (MMHP) model under the assumption that $S=2$, and utilizes a variational approximation to overcome computational challenges. When the underlying Markov process is in the \textit{active} state ($Z(t) = 1$), the events occur in bursty patterns as in a Hawkes process with intensity $\lambda_1(t)= \lambda_1 + \sum_{t_m<t}\alpha e^{-\beta(t-t_m)}$, while in the \textit{inactive} state ($Z(t) = 0$), the dynamics switch to a quieter period following a homogeneous Poisson process with constant rate $\lambda_0$. The parameter set for a MMHP is then $\Theta:=\{\lambda_0,\  \lambda_1,\ \alpha,\ \beta,\ Q\}$.

\subsection{Network point processes} 
Consider a network consisting of a fixed set of $N$ nodes, $V=\{1,2,...,N\}$. The observation of event arrival times on a network is defined on $\mathbb{R}\times V\times V$, where $N(t,i,j):= N^{i,j}(t)$ is the number of interactions between node $i$ and node $j$ during the time interval $(0,t]$. This is essentially a marked point process with finite mark space and is also called a multivariate point process \citep{cox1972multivariate}. Throughout this paper, we will consider a directed network, hence the sequence $i,j$ conveys information that $i$ is the sender and $j$ is the receiver. We also assume that there are no self-loops, i.e., $i\neq j$. Define the history of interactions up to time $T$ as $\mathcal{H}_V(T)=\{(t_1,i_1,j_1),...,(t_M,i_M,j_M)\}$, where $M$ is the total number of events on the network. $\mathcal{H}_V(T)$ can also be represented by $\bigcup_{i,j \in V, i\neq j} \mathcal{H}^{i,j}(T)$, where $\mathcal{H}^{i,j}(T)$ is the history for pair $(i,j)$ containing $M^{i,j}$ event times, i.e.$\{t^{i,j}_m, m=1,2,...,M^{i,j}\}$.

The conditional intensity for the marginal counting process $N^{i,j}(t)$ between a pair $(i,j)$ is defined as the instantaneous expected rate of events occurring around a time $t$ given the history:
$$\lambda^{i,j}(t|\mathcal{H}_V(t))=\lim_{\Delta t\to 0} \frac{Pr([N^{i,j}(t+\Delta t)-N^{i,j}(t)]=1|\mathcal{H}_V(t))}{\Delta t}.$$

Following proposition 7.3.\Romannum{3} in \cite{daley2003introduction}, the likelihood function is given by $$\prod_{i=1}^N\prod_{j\neq i}^N\Big[\prod_{m=1}^{M^{i, j}}\lambda^{i,j}(t_m^{i, j}|\mathcal{H}_V(t_m^{i, j}))\Big]\exp\Big\{-\int_0^T\lambda^{i,j}(s|\mathcal{H}_V(s))ds\Big\}.$$


The simplest model for point processes on a network is a {\em homogeneous} network point process model, where $\lambda^{i,j}(t_m^{i, j}|\mathcal{H}_V(t_m^{i, j}))$ does not depend on $i,j$. As dynamic interactions on a social network are known to be heterogeneous, clustered, and structured by the underlying social distance among actors, a few models have been developed recently to take into account network heterogeneity and structure. However, the discussion on model assessment and diagnosis remains limited in the literature. To adjust for higher-degree actors in a network, and pair-specific covariates, \cite{perry2013point} proposes a multivariate point process with the model assumption that the intensities between pairs are decided by a function of a sender-specific baseline rates. By treating the sender baseline intensity rates as nuisance parameters, the model estimation is carried out by maximizing the log-partial-likelihood. For model checking, they calculate and visualize normalized residuals. These residuals do not account for the sender-specific baseline rates and are not used to examine how well the proposed model explains network dependence in event arrivals among pairs of actors. \cite{zipkin2016point} models the arrival times of interaction events between pairs of actors of known identity using independent Hawkes processes and uses the fitted pairwise processes to {\em resolve} the actors' identities of an interaction event given only the event's arrival time and no actor identities. They validate the fitted model by comparing the resolved actor identities for a hold-out set of event times against the ground truth. Both of \cite{linderman2014discovering} and \cite{yang2017decoupling} assume a latent network structure and consider a reciprocity effect, using a Hawkes process model. The models are validated by prediction of hold-out links and interpretation of the latent space structure. The above validation procedures are inadequate to inspect the proposed models' lack-of-fit in terms of temporal trends and/or network structure. We will show, in Sections~\ref{sec:time} and~\ref{sec:network}, a systematic approach for model checking and diagnostics, with theoretically well-grounded diagnostic statistics and visualization tools.



\section{Model checking for time heterogeneity}
\label{sec:time}

We first consider univariate point processes. The goal is to check that the fitted model can capture the variability in temporal trends. \cite{zheng_markov-modulated_2019} shows that the \textit{Markov-modulated Hawkes process} (MMHP) can model the patterns of event dynamics that are sporadic with bursts and long wait times. In our simulated case studies, we will generate event times according to a MMHP, and compare the model fit of a homogeneous point process model, a Hawkes process model, a MMPP model and a MMHPSD model. In our real data example, we fit the above five models to the interaction events between one pair of mice as in Figure~\ref{fig:real_example}-(a).

\subsection{Time rescaling theorem}
\label{sec:time_1}

One approach to test the goodness-of-fit of point process models is to apply the time rescaling theorem \citep{brown2002time}. 
It states that if $\{t_m\}_{m=1:M}$ is a realization of events from a point process with conditional intensity $\lambda(t|\mathcal{H}(t))$, then the rescaling transformation $\int_0^{t_m}\lambda(s|\mathcal{H}(s))ds$ over $m=1,...,M$ yields a Poisson process with rate 1. Hence, the \textit{rescaled-inter-event times}, defined as $\{\Lambda_m:=\int_{t_{m-1}}^{t_m}\lambda(s|\mathcal{H}(s))ds\}_{m=1:M}$ are independently and identically distributed as exponential random variables with rate 1.

\begin{figure}[h!]
 \centering
 \includegraphics[width=\textwidth]{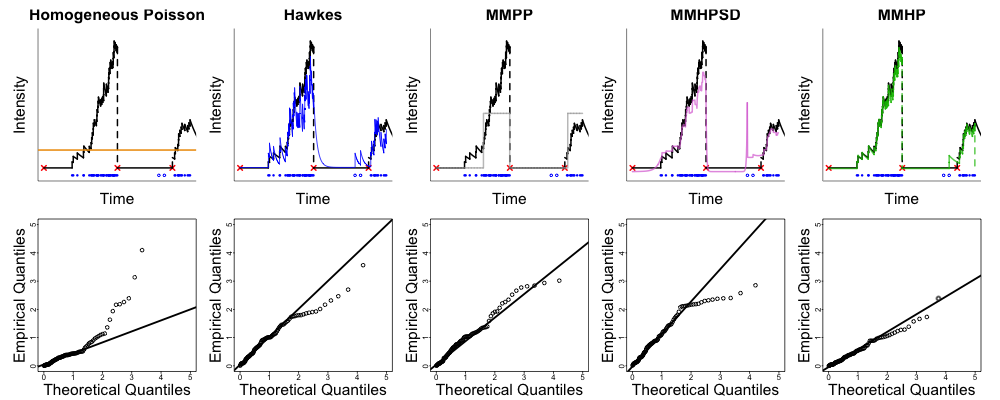}
 \caption{Intensity estimation and Q-Q plots for testing the distribution of \textit{rescaled-inter-event times} after fitting the five models to synthetic data. Upper panels plot a comparison between intensity functions versus the true intensity, where the blue dots are events, black lines indicate the true intensity function and colored lines are the estimations. Knowing the true intensity, the red crosses are state transitions. Lower panels give Q-Q plots for testing whether the \textit{rescaled-inter-event times} are distributed as exponential random variables with rate 1.}
 \label{fig:sim_compensator}
\end{figure}

\textbf{Case study \Romannum{1}: simulated examples.} We simulate a MMHP using thinning algorithm \citep{lewis1979simulation} with parameter value $\Theta=\{\lambda_0=1,\  \lambda_1=1.1,\ \alpha=1.6,\ \beta=1.9,\ Q = \bigl( \begin{smallmatrix}-0.2 & 0.2\\ 0.4 & -0.4\end{smallmatrix}\bigr)\}$ and termination time $T=100$. We fit the five models to the simulated data separately. Figure \ref{fig:sim_compensator} shows the estimated intensity versus the true intensity, and Q-Q plots to test for goodness-of-fit. The MMHP model can recover the true intensity precisely and reveals a nearly exponential distributed \textit{rescaled-inter-event times}, i.e., $\{\Lambda_m\}_{m=1:M}$. The Q-Q plot for the homogeneous Poisson process model jumps nearly vertically at larger quantiles, since it fails to capture the intensity at state 0 and thus has a heavier tail than the exponential distribution. For the Hawkes process and MMHPSD models, the Q-Q plots show that the empirical distribution has a spike concentrated on a larger value than the theoretical median. Both models fail to capture the high intensity during state 1 because they compromise their parameter estimation between state 1 and state 0. The MMPP model yields an empirical distribution with a lighter right tail and is not flexible enough to capture the intensity during state 1. 

\begin{figure}[h!]
 \centering
 \includegraphics[width=\textwidth]{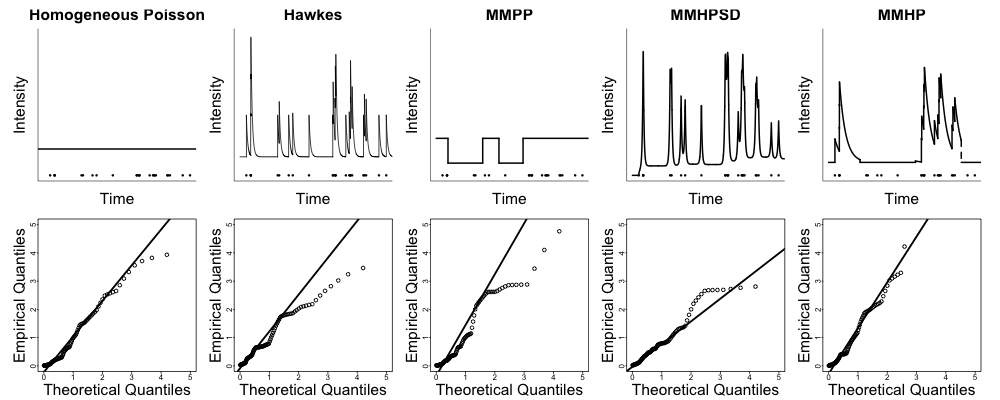}
 \caption{Model fit diagnostics for \citet{williamson2016temporal} data: intensity estimation and Q-Q plots after fitting the five models to real event time data between one pair of mice as shown in Figure~\ref{fig:real_example}-(a). Upper panels plot estimated intensity functions after fitting the five models, which are indicated by the black lines. The black dots are the events occurring over time. Lower panels are Q-Q plots to test goodness-of-fit.}
 \label{fig:real_compensator}
\end{figure}

\textbf{Case study \Romannum{2}: an application to \citet{williamson2016temporal} data.} Figure \ref{fig:real_compensator} shows the intensity estimation and Q-Q plots after fitting the five models to observed event times for the pair of mice shown in Figure~\ref{fig:real_example}-(a). MMHP fits the data most reasonably according to the Q-Q plot, whereas the other four models show some indications of lack-of-fit. The fitted homogeneous Poisson process, Hawkes process and MMPP models all show a lighter right tail than the exponential distribution, while the fitted MMHPSD model has a heavier right tail. By using Q-Q plots to examine the distribution of $\{\Lambda_m\}_{m=1:M}$, we can compare model fit across competing point process models and can conclude that the MMHP model captures the sporadic and bursty event dynamics of mice social interactions.

\subsection{Residual process}

According to the Doob-Meyer decomposition theorem \citep{andersen2012statistical}, given a counting process $N(t)$ with its conditional intensity function $\lambda(t|\mathcal{H}(t))$, the \textit{residual process} $M(t)$, defined as $M(t)=N(t)-\int_0^t\lambda(s|\mathcal{H}(s))ds$, is a martingale. Hence, when a point process model is fitted to data and gives an estimated intensity function $\hat{\lambda}(t)$, the \textit{raw residual process}
$$R(t)=N(t)-\int_0^t\hat{\lambda}(s)ds$$ can be used to inspect the fit of the model by measuring its discrepancy from 0. Here, we evaluate the raw residual process at the final time point, $T$.

The variance of the above residual process $M(T)$ depends on the intensity function, $\mbox{var}(M(t))=\int_0^t\lambda(s)ds$. To compare model fit between scenarios with different intensity functions, it is desirable to compute a {\em standardized} residual \citep{baddeley2005residual}, similar to the Pearson residuals for linear regression. \citet{clements2011residual} proposes the \textit{Pearson residual process} for a counting process as an intensity-weighted version of the raw residual process, $$\mbox{PR}(t)=\sum_{t_m<t}\frac{1}{\sqrt{\lambda(t_m)}}-\int_0^t\sqrt{\lambda(s)}ds.$$ Under the true model specification, the variance of $PR(t)$ does not depend on the intensity, i.e., $\mbox{var}(\mbox{PR}(t))=t$ with mean 0 \citep{clements2011residual}. 

\begin{figure}[h!]
 \centering
 \includegraphics[width=0.9\textwidth]{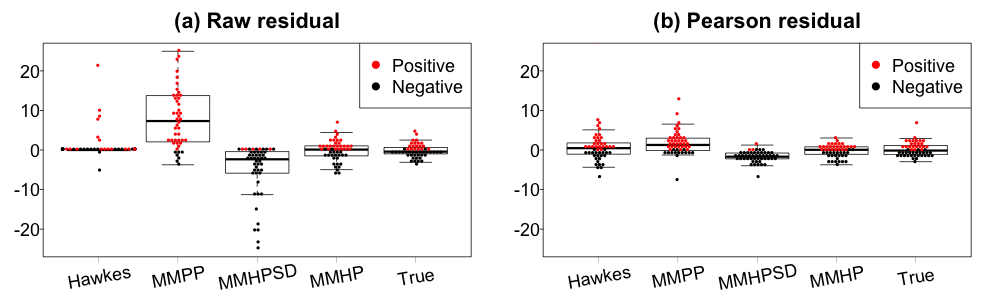}
 \caption{Beeswarm boxplots for raw residual and Pearson residual processes after fitting the four models to synthetic data from 50 independent MMHPs. The residuals calculated from the true intensities are also plotted for comparison.}
 \label{fig:sim_residual}
\end{figure}

\textbf{Case study \Romannum{1}: simulated examples.} We simulate 50 independent MMHPs with the same parameter settings as in Section \ref{sec:time_1}. After fitting Hawkes process, MMPP, MMHPSD and MMHP models to the data, the raw residual $R(T)$ and Pearson residual $\mbox{PR}(T)$ are calculated. Due to the formulation of the Pearson residual process, the integral part cannot be calculated explicitly, and hence, is approximated by numerical integration. Boxplots for 50 raw residuals and Pearson residuals are shown in Figure \ref{fig:sim_residual}-(a) and (b) separately. We can see that the Hawkes process and MMPP models yield positive residuals most of the time, which means that the models tend to underestimate the intensity. The Hawkes process biased its parameter estimation in order to compromise to the low intensity in state 0, and underestimated the intensity in state 1. The MMPP model cannot capture the volatility of the intensity function in state 1 and yield an underestimated intensity. On the contrary, the MMHPSD model overestimates the intensity, mostly due to its over-sensitive intensity function with the step-wise kernel. 

\begin{figure}[t]
 \centering
 \includegraphics[width=0.9\textwidth]{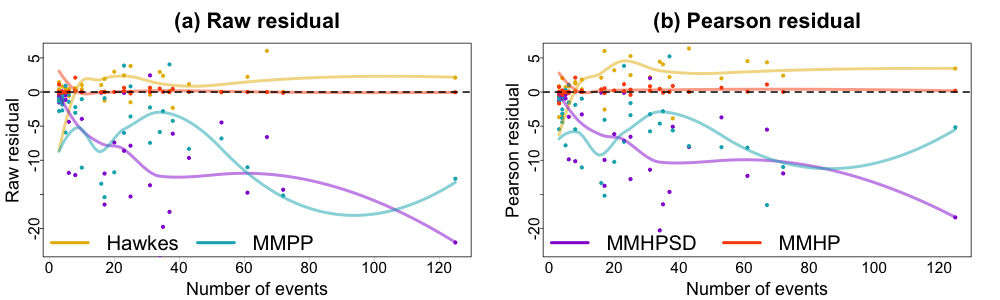}
 \caption{Raw residuals and Pearson residuals after fitting the four models to all pairs in the cohort of mice as in Figure~\ref{fig:real_example}-(b) from \citet{williamson2016temporal}. Each model is represented by a different color, with residual values (dots) and a smoothed LOWESS line.}
 \label{fig:real_residual} 
\end{figure}

\textbf{Case study \Romannum{2}: an application to \citet{williamson2016temporal} data.} Using the data from \citet{williamson2016temporal}, we fit the above four models to all pairs of mice in a cohort separately, and then evaluated the raw residuals and Pearson residuals at final observation time $T$ for all pairs. Figure \ref{fig:real_residual} plots the residuals versus the number of events that occurred between a pair, with smoothing regression lines using LOWESS method \citep{cleveland1979robust}. The MMHP model has residuals closest to 0 and is not sensitive to the number of events (in a sense, the number of events is a rough indicator of intensity) while the other models fail. Comparing the raw residual of the MMPP and MMHPSD models with their Pearson residual, we also observe that the raw residuals have more variability over the number of events. This suggests that the Pearson residual is more appropriate in the network setting, since it is important to have a measurement that is comparable across pairs with varying numbers of pairwise interactions.

\section{Model checking for network heterogeneity and structure}
\label{sec:network}

In this section, we provide tools for assessing network point process models. We will focus on MMHP in a network setting and define the marginal intensity for a pair $(i,j)$ as 
\begin{equation}
\label{eq:network_mmhp}
\lambda^{i,j}(t|\mathcal{H}_V(t))=
\begin{cases}
\lambda_1^{i,j}+\sum_{t_m^{i,j}<t}\alpha^{i,j}e^{-\beta^{i,j}(t-t_m^{i,j})},& Z^{i,j}(t)=1 \\
\lambda_0^{i,j},& Z^{i,j}(t)=0,
\end{cases}
\end{equation}
where $Z^{i,j}(t)$ is an independent CTMC across all pairs $(i,j)$. We denote the parameters for $(i,j)$ as $\Theta^{i,j}=\{\lambda_0^{i,j},\  \lambda_1^{i,j},\ \alpha^{i,j},\ \beta^{i,j},\ Q^{i,j}\}$. 

\subsection{Kolmogorov-Smirnov test}
\label{sec:network_ks}

For each pair $(i,j)$, we want to utilize the time rescaling theorem in Section \ref{sec:time_1} and test whether the \textit{rescaled-inter-event times}, $\{\Lambda^{i,j}_m\}_{m=1:M^{i,j}}$ distributed as exponential random variables with rate 1, where $\Lambda^{i,j}_m=\int_{t^{i,j}_{m-1}}^{t^{i,j}_m}\lambda^{i,j}(s|\mathcal{H}_V(s))ds$. To summarize the model fit across pairs in the network, it is important to quantify the test results with a unified measurement. We use the Kolmogorov-Smirnov statistics, $\mbox{KS}^{i,j}=\sup_x|\hat{F}^{i,j}(x)-F(x)|$ for each pair $(i,j)$, where $\hat{F}^{i,j}(x)$ is the empirical distribution of the \textit{rescaled-inter-event times} and $F(x)$ is the cumulative distribution function of the exponential distribution. Larger values of the K-S statistics indicate a larger discrepancy between the model and the network point pattern data.
 
\begin{figure}[t]
 \subfigure
 {
 \includegraphics[width=0.245\linewidth]{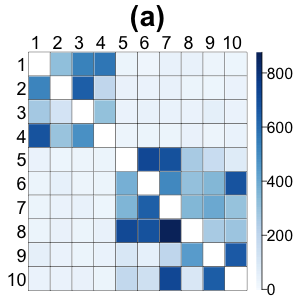}
 }
 \subfigure
 {
 \includegraphics[width=0.208\linewidth]{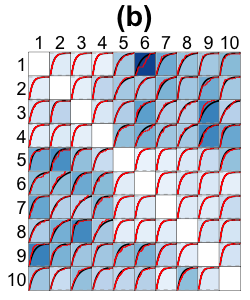}
 }
 \subfigure
 {
 \includegraphics[width=0.208\linewidth]{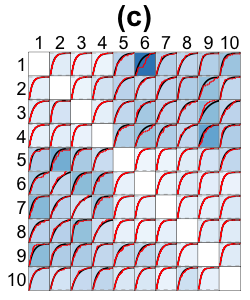}
 }
 \subfigure
 {
 \includegraphics[width=0.245\linewidth]{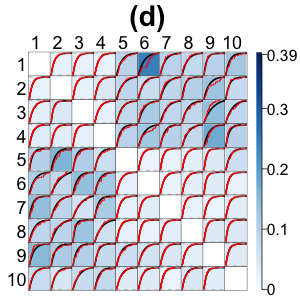}
 }
 \caption{(a) A heatmap of simulated network point process data: the total number of events occurred between each pair. (b)(c) Matrices of K-S statistics after fitting \textit{homogeneous} and \textit{block}-network models to the simulated data in (a). The rows and columns of each matrix correspond to senders and receivers, respectively. Color shades reflect the values of the K-S test statistics. (d) Matrix of K-S statistics calculated from true intensity function.}
 \label{fig:sim_all_ks}
\end{figure}

\textbf{Case study \Romannum{1}: simultated examples.} We simulate network point process data with 10 nodes until time $T=500$ as follows: for each pair $(i,j)$, the marginal intensity $\lambda^{i,j}(s|\mathcal{H}_V(s))$ follows (\ref{eq:network_mmhp}). All the pairs $(i,j)$ share the same paramters that $\lambda_0^{i,j}=0.05,\ \lambda_1^{i,j}=0.08,\ \beta^{i,j}=22,\ Q^{i,j} = \bigl( \begin{smallmatrix}-0.01 & 0.01\\ 0.04 & -0.04\end{smallmatrix}\bigr)$. The matrix $\boldsymbol{\alpha}=[\alpha^{i,j}]_{N\times N}$ is block-structured,
\[
\alpha_{ij}=
\begin{cases}
20, & i,j\in B_k \\
0.5, & i\in B_k, j\in B_l, k\neq l
\end{cases}.
\]
where class $B_1=\{1,2,3,4\}$ and $B_2=\{5,6,...,10\}$. The total number of events between pair $(i,j)$ until time $T$, $N^{i,j}(T)$ is shown in Figure \ref{fig:sim_all_ks}-(a), which has a clear block structure.

We compare the fit of the following two models: (1) across all pairs, the parameter set $\Theta^{i,j}$ shares a same set of values (referred as \textit{homogeneous}-network model); (2) $\alpha^{i,j}$ varies across pairs such that $\boldsymbol{\alpha}$ has a block structure with the same classes as in the simulation (all other elements of $\Theta^{i,j}$ are the same across all pairs, as in the first model). This is referred to as \textit{block}-network model.

Figure~\ref{fig:sim_all_ks}-(b) shows the K-S statistics over the network after fitting the \textit{homogeneous}-network model to the simulated data, whereas (c) is for the \textit{block}-network model. Figure~\ref{fig:sim_all_ks}-(d) displays results from the true intensity function, which has the lowest K-S statistics. \textit{Homogeneous}-network model (Figure \ref{fig:sim_all_ks} - (b)) fits the data the worst, since it has higher K-S statistics values across network compared to the other models. Figure \ref{fig:sim_all_ks} - (b) also shows a block-structured K-S statistics and demonstrates that the \textit{homogeneous}-network model performs worse for the between-block pairs, since within-block pairs generate more events and contribute more to the likelihood function.

\begin{figure}[h]
 \centering
   \begin{minipage}[c]{0.57\textwidth}
    \includegraphics[width=\linewidth]{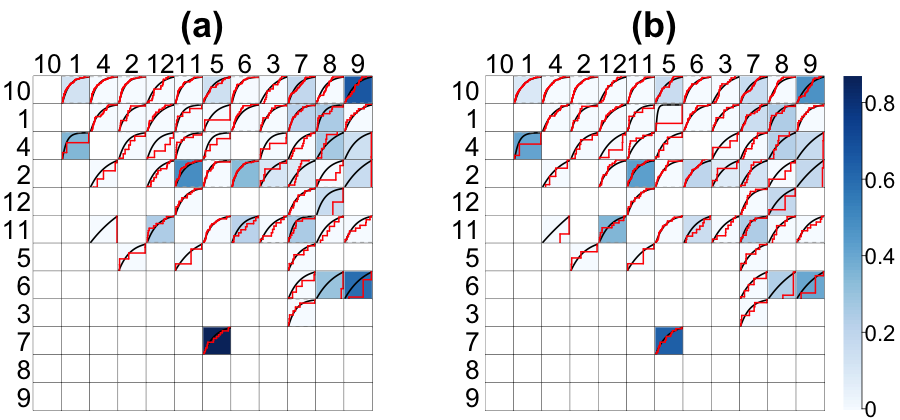}
  \end{minipage}\hfill
  \begin{minipage}[c]{0.4\textwidth}
    \vspace{20pt}
    \caption{Matrices of K-S statistics after fitting \textit{homogeneous}-network model (in (a)) and \textit{heterogeneous}-network model (in (b)) to the mice interaction data. Color shades reflects the values of the K-S test statistics.}  \label{fig:real_all_ks}
  \end{minipage}
  \vspace{-8pt}
\end{figure}

\textbf{Case study \Romannum{2}: an application to \citet{williamson2016temporal} data.} For the interaction event times among a group of mice as shown in Figure \ref{fig:real_example}-(b), we fit the following two models: (1) the parameter set $\Theta^{i,j}$ for MMHP is the same across pairs (referred as \textit{homogeneous}-network model); (2) $\Theta^{i,j}$ is allowed to vary freely across pairs (referred as \textit{heterogeneous}-network model). Figure \ref{fig:real_all_ks} plots the matrices of K-S statistics after fitting the above two models. Since the \textit{heterogeneous}-network model has more flexibility across pairs and can adapt well on various pairs' dynamics, its K-S statistics are smaller and the model fits the data better. Especially for the pair $(7,5)$, they started engaging in interactions during the last period of observations, which sets them apart from the other pairs, as shown in \ref{fig:real_example}-(b). The \textit{homogeneous}-network model fits this pair poorly, whereas the \textit{heterogeneous}-network model better captures this pattern.
\vspace{-20pt}
\vspace{-10pt}
\subsection{Structure score based on Pearson residual matrix}

The Pearson residual is more valuable for diagnosing model fit in the network setting, since it is comparable across pairs and not influenced by intensity function. For a pair $(i,j)$, the Pearson residual is 

$$\mbox{PR}_{i,j}(T)=\sum_{t^{i,j}_m}\frac{1}{\sqrt{\hat{\lambda}^{i,j}(t^{i,j}_m)}}-\int_0^T\sqrt{\hat{\lambda}^{i,j}(s)}ds.$$

To assess model fit, we propose to quantify the network structure in the Pearson residual matrix, $\mbox{PR}:=[\mbox{PR}_{i,j}]_{N\times N}$. In order to spot systematic overestimation or underestimation issues, we separate the Pearson residual matrix into two matrices for the positive and negative residual values respectively. We preserve the positive residuals in the {\em underestimation} matrix $\mbox{PR}^+$, such that $\mbox{PR}^+_{i,j}=\mbox{PR}_{i,j}$ if $\mbox{PR}_{i,j}>0$ and $\mbox{PR}^+_{i,j}=0$ otherwise. Similarly, we create the {\em overestimation} matrix $\mbox{PR}^-$ using the absolute values of negative residuals. We conduct nonnegative matrix factorization (NMF) \citep{lee2001algorithms} on each of the matrices, $A\approx WH$, where $A\in\mathbb{R}^{N\times N}$ is $\mbox{PR}^+$ or $\mbox{PR}^-$ and $W\in\mathbb{R}^{N\times K}, H\in\mathbb{R}^{K\times N}$. $K$ is usually set to be much smaller than $N$, so that $WH$ is a lower-rank approximation of $A$. We introduce a {\em matrix structure score} as $\frac{||A-WH||_F}{||A||_F}$, where $||\cdot||_F$ is the matrix Frobenius norm. If NMF can recover the residual matrix well, the score is larger, suggesting that the model fits the data worse and yields a more structured residual matrix.

\begin{figure}[t]
 \centering
 \includegraphics[width=\textwidth]{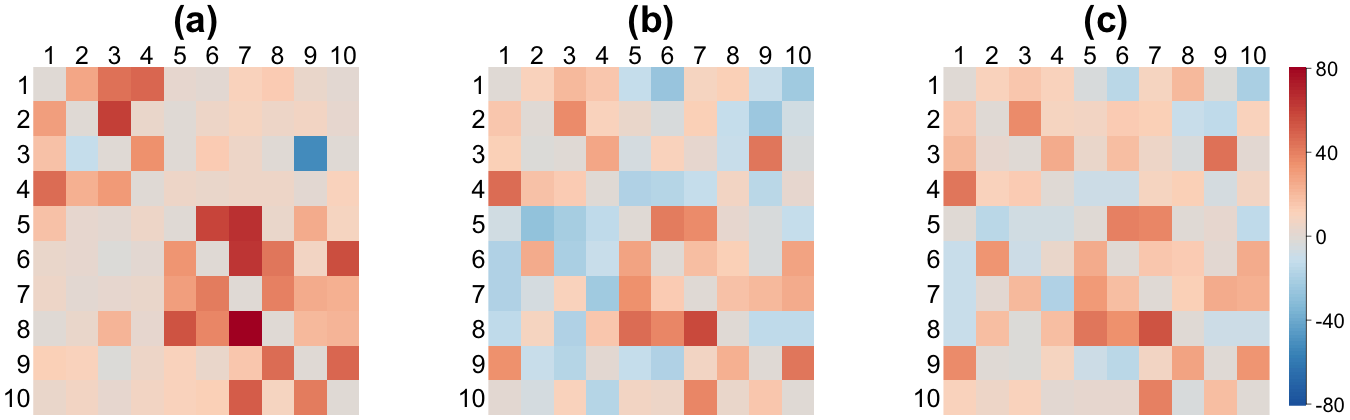}
 \caption{Matrices of the Pearson residuals after fitting models to simulated data. (a)(b) are results after fitting the \textit{homogeneous} and \textit{block}-network models to the simulated data. (c) is from the true model that is used to generate simulation. Red colors represent positive residuals, indicating that the model underestimates the intensity function, whereas blue colors represent negative residuals and hence overestimation. }
 \label{fig:sim_pearson}
\end{figure}

\textbf{Case study \Romannum{1}: simulated examples.} Using simulated data as in Section \ref{sec:network_ks}, we calculate the Pearson residual matrix after fitting the \textit{homogeneous} and \textit{block}-network models and plot them in Figure \ref{fig:sim_pearson}-(a) and (b) respectively. For a better comparison, we also use the true model intensity to calculate the residuals and plot in Figure \ref{fig:sim_pearson}-(c). Table \ref{table:simulation} compares the matrix structure score of the underestimation $\mbox{PR}^+$ and overestimation matrix $\mbox{PR}^-$ across models by using $K=2$. The \textit{homogeneous} model exhibits larger positive residuals within-block. This demonstrates that the estimation of the intensities for within-block pairs are biased towards lower values because the model is network-homogeneous and not adequate to capture intense interactions of within-block pairs. The matrix structure score of this model is higher in both overestimation and underestimation matrices as shown in Table \ref{table:simulation}, which is a quantified validation for its lack-of-fit.

\begin{center}
\begin{tabular}{cc}
\centering
    \begin{minipage}{0.42\linewidth}

        \begin{tabular}{ccccc}\hline
         \ & homogeneous \ & block \ & true \ \\\hline
positive \ & 0.46 \ & 0.41  \ & 0.41 \ \\
negative \ & 0.93 \ & 0.43  \ & 0.36 \ \\\hline
        \end{tabular}

    \captionof{table}{Structure scores for simulated examples.}
    \label{table:simulation}
    \end{minipage} &
    \qquad
    \begin{minipage}{.5\linewidth}
    
        \begin{tabular}{ccccc}\hline
         \ & homogeneous \ & heterogeneous \ \\\hline
positive \ & 0.79 \ & 0.69      \ \\
negative \ & 0.77 \ & 0.55      \ \\\hline
        \end{tabular}
    \captionof{table}{Structure scores for the \citet{williamson2016temporal} example.}
    \label{table:real} 
    \end{minipage} 
\end{tabular}
\vspace{4pt}
\end{center}

\begin{figure}[h!]
 \centering
 \includegraphics[width=0.68\textwidth]{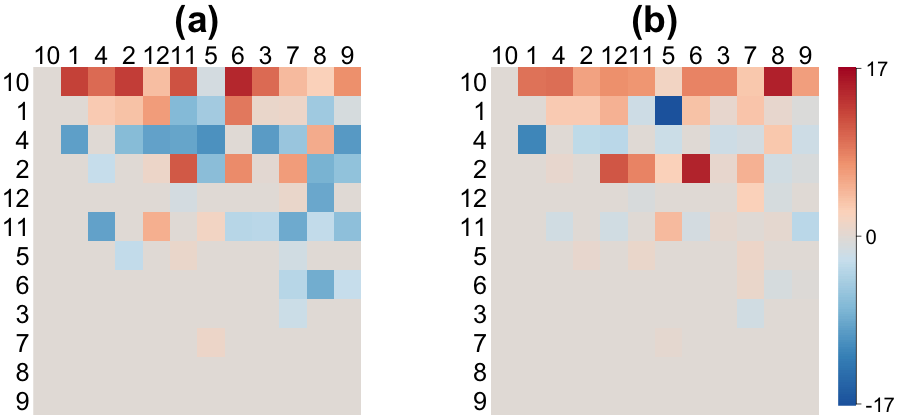}
 \caption{Matrices of the Pearson residuals after fitting models to \citet{williamson2016temporal} data. (a) is for \textit{homogeneous}-network model and (b) is for \textit{heterogeneous}-network model. The rows and columns of all matrices are reordered by the I\&SI dominance rank.}
 \label{fig:real_pearson}
\end{figure}

\vspace{4pt} \textbf{Case study \Romannum{2}: an application to \citet{williamson2016temporal} data.} Figure \ref{fig:real_pearson} shows the Pearson residual matrix after fitting \textit{homogeneous} and \textit{heterogeneous}-network models to the mice interaction data. The models' assumptions are defined in Section \ref{sec:network_ks}. The \textit{homogeneous}-network model tends to overestimate the intensity function when individual 4 and 11 are the senders, because the inference is biased by a large number of interactions initiated by active individuals. Thus, this pattern in the residual matrix yields a larger matrix structure score, especially for the negative residuals, as shown in Table \ref{table:real}, where $K$ is set to be 2. It suggests that the \textit{homogeneous}-network model is not adequate to capture the network dynamics, while the \textit{heterogeneous}-network model has more flexibility and is superior in terms of Pearson residual and structure score.

\section{Summary}

In this paper, we propose diagnostic statistics and visualization tools for network point process models. The evaluation techniques are theoretically well-grounded and can be applied to intensity-based models for network point pattern data generally. We use simulations and real data examples to demonstrate the utility of our approaches. By inspecting lack-of-fit in terms of both temporal dependence and network structures, the proposed suite of diagnostic statistics and visualizations can reveal deficiencies in network point process models and provide important insights that can lead to model improvements.

\section{Acknowledgment}

This material is based on research sponsored by DARPA agreement number D17AC00001. The content of the information does not necessarily reflect the position or the policy of the Government, and no official endorsement should be inferred.

\vspace{20pt}
{
\renewcommand{\bibname}{References\vspace*{-30pt}}
\renewcommand{\clearpage}{} 
\bibliographystyle{abbrvnat}
\bibliography{diagnosis}

\begin{thebibliography}{23}
\providecommand{\natexlab}[1]{#1}
\providecommand{\url}[1]{\texttt{#1}}
\expandafter\ifx\csname urlstyle\endcsname\relax
  \providecommand{\doi}[1]{doi: #1}\else
  \providecommand{\doi}{doi: \begingroup \urlstyle{rm}\Url}\fi

\bibitem[Andersen et~al.(2012)Andersen, Borgan, Gill, and
  Keiding]{andersen2012statistical}
P.~K. Andersen, O.~Borgan, R.~D. Gill, and N.~Keiding.
\newblock \emph{Statistical models based on counting processes}.
\newblock Springer Science \& Business Media, 2012.

\bibitem[Baddeley et~al.(2005)Baddeley, Turner, M{\o}ller, and
  Hazelton]{baddeley2005residual}
A.~Baddeley, R.~Turner, J.~M{\o}ller, and M.~Hazelton.
\newblock Residual analysis for spatial point processes (with discussion).
\newblock \emph{Journal of the Royal Statistical Society: Series B (Statistical
  Methodology)}, 67\penalty0 (5):\penalty0 617--666, 2005.

\bibitem[Barabasi(2005)]{barabasi2005origin}
A.-L. Barabasi.
\newblock The origin of bursts and heavy tails in human dynamics.
\newblock \emph{Nature}, 435\penalty0 (7039):\penalty0 207--211, 2005.

\bibitem[Brown et~al.(2002)Brown, Barbieri, Ventura, Kass, and
  Frank]{brown2002time}
E.~N. Brown, R.~Barbieri, V.~Ventura, R.~E. Kass, and L.~M. Frank.
\newblock The time-rescaling theorem and its application to neural spike train
  data analysis.
\newblock \emph{Neural computation}, 14\penalty0 (2):\penalty0 325--346, 2002.

\bibitem[Clements et~al.(2011)Clements, Schoenberg, and
  Schorlemmer]{clements2011residual}
R.~A. Clements, F.~P. Schoenberg, and D.~Schorlemmer.
\newblock Residual analysis methods for space-time point processes with
  applications to earthquake forecast models in california.
\newblock \emph{The Annals of Applied Statistics}, pages 2549--2571, 2011.

\bibitem[Cleveland(1979)]{cleveland1979robust}
W.~S. Cleveland.
\newblock Robust locally weighted regression and smoothing scatterplots.
\newblock \emph{Journal of the American statistical association}, 74\penalty0
  (368):\penalty0 829--836, 1979.

\bibitem[Cox and Lewis(1972)]{cox1972multivariate}
D.~R. Cox and P.~A.~W. Lewis.
\newblock Multivariate point processes.
\newblock In \emph{Proc. 6th Berkeley Symp. Math. Statist. Prob}, volume~3,
  pages 401--448, 1972.

\bibitem[Daley and Vere-Jones(2003)]{daley2003introduction}
D.~J. Daley and D.~Vere-Jones.
\newblock \emph{An introduction to the theory of point processes: volume I:
  elementary theory and methods}.
\newblock Springer Science \& Business Media, 2003.

\bibitem[Fan and Shelton(2009)]{fan2009learning}
Y.~Fan and C.~R. Shelton.
\newblock Learning continuous-time social network dynamics.
\newblock In \emph{Proceedings of the Twenty-Fifth Conference on Uncertainty in
  Artificial Intelligence}, pages 161--168. AUAI Press, 2009.

\bibitem[Fischer and Meier-Hellstern(1993)]{fischer1993markov}
W.~Fischer and K.~Meier-Hellstern.
\newblock The {M}arkov-modulated {P}oisson process ({MMPP}) cookbook.
\newblock \emph{Performance evaluation}, 18\penalty0 (2):\penalty0 149--171,
  1993.

\bibitem[Hawkes(1971)]{hawkes1971spectra}
A.~G. Hawkes.
\newblock Spectra of some self-exciting and mutually exciting point processes.
\newblock \emph{Biometrika}, pages 83--90, 1971.

\bibitem[Lee and Seung(2001)]{lee2001algorithms}
D.~D. Lee and H.~S. Seung.
\newblock Algorithms for non-negative matrix factorization.
\newblock In \emph{Advances in neural information processing systems}, pages
  556--562, 2001.

\bibitem[Lewis and Shedler(1979)]{lewis1979simulation}
P.~W. Lewis and G.~S. Shedler.
\newblock Simulation of nonhomogeneous poisson processes by thinning.
\newblock \emph{Naval research logistics quarterly}, 26\penalty0 (3):\penalty0
  403--413, 1979.

\bibitem[Linderman and Adams(2014)]{linderman2014discovering}
S.~Linderman and R.~Adams.
\newblock Discovering latent network structure in point process data.
\newblock In \emph{International Conference on Machine Learning}, pages
  1413--1421, 2014.

\bibitem[Perry and Wolfe(2013)]{perry2013point}
P.~O. Perry and P.~J. Wolfe.
\newblock Point process modelling for directed interaction networks.
\newblock \emph{Journal of the Royal Statistical Society: Series B (Statistical
  Methodology)}, 75\penalty0 (5):\penalty0 821--849, 2013.

\bibitem[Rabiner(1989)]{rabiner1989tutorial}
L.~R. Rabiner.
\newblock A tutorial on hidden markov models and selected applications in
  speech recognition.
\newblock \emph{Proceedings of the IEEE}, 77\penalty0 (2):\penalty0 257--286,
  1989.

\bibitem[Saito et~al.(2009)Saito, Kimura, Ohara, and Motoda]{saito2009learning}
K.~Saito, M.~Kimura, K.~Ohara, and H.~Motoda.
\newblock Learning continuous-time information diffusion model for social
  behavioral data analysis.
\newblock In \emph{Asian Conference on Machine Learning}, pages 322--337.
  Springer, 2009.

\bibitem[Schmid and de~Vries(2013)]{schmid2013finding}
V.~S. Schmid and H.~de~Vries.
\newblock Finding a dominance order most consistent with a linear hierarchy: an
  improved algorithm for the i\&si method.
\newblock \emph{Animal Behaviour}, 86\penalty0 (5):\penalty0 1097--1105, 2013.

\bibitem[Wang et~al.(2012)Wang, Bebbington, and Harte]{wang2012markov}
T.~Wang, M.~Bebbington, and D.~Harte.
\newblock Markov-modulated {H}awkes process with stepwise decay.
\newblock \emph{Annals of the Institute of Statistical Mathematics},
  64\penalty0 (3):\penalty0 521--544, 2012.

\bibitem[Williamson et~al.(2016)Williamson, Lee, and
  Curley]{williamson2016temporal}
C.~M. Williamson, W.~Lee, and J.~P. Curley.
\newblock Temporal dynamics of social hierarchy formation and maintenance in
  male mice.
\newblock \emph{Animal Behaviour}, 115:\penalty0 259--272, 2016.

\bibitem[Wu et~al.(2019)Wu, Zheng, and Curley]{zheng_markov-modulated_2019}
J.~Wu, T.~Zheng, and J.~Curley.
\newblock Markov-modulated {H}awkes processes for sporadic and bursty event
  occurrences.
\newblock \emph{arXiv: 1903.03223}, 2019.

\bibitem[Yang et~al.(2017)Yang, Rao, and Neville]{yang2017decoupling}
J.~Yang, V.~Rao, and J.~Neville.
\newblock Decoupling homophily and reciprocity with latent space network
  models.
\newblock In \emph{UAI}, 2017.

\bibitem[Zipkin et~al.(2016)Zipkin, Schoenberg, Coronges, and
  Bertozzi]{zipkin2016point}
J.~R. Zipkin, F.~P. Schoenberg, K.~Coronges, and A.~L. Bertozzi.
\newblock Point-process models of social network interactions: Parameter
  estimation and missing data recovery.
\newblock \emph{European Journal of Applied Mathematics}, 27\penalty0
  (3):\penalty0 502--529, 2016.

\end{thebibliography}
}
\end{document}